\documentclass[twocolumn,showpacs,preprintnumbers,amsmath,amssymb]{revtex4}
\usepackage[dvips]{graphicx}
\usepackage{color}
\usepackage{amssymb}
\usepackage{amsmath}
\usepackage{units}
\usepackage[ansinew]{inputenc} % to allow German characters as {\"a} to be typeset correctly
\usepackage{graphicx}% Include figure files

\pdfoutput=1

\newcommand{\Py}{Ni$_{81}$Fe$_{19}$}

\begin{document}

\title{A spin-wave frequency doubler by domain wall oscillation}

\author{Sebastian~J.~Hermsdoerfer}
\affiliation{Fachbereich Physik and Landesforschungszentrum
OPTIMAS, Technische Universit\"at Kaiserslautern,
Erwin-Schr\"odinger-Stra{\ss}e 56, 67663 Kaiserslautern, Germany}

\author{Helmut~Schultheiss}
\affiliation{Fachbereich Physik and Landesforschungszentrum
OPTIMAS, Technische Universit\"at Kaiserslautern,
Erwin-Schr\"odinger-Stra{\ss}e 56, 67663 Kaiserslautern, Germany}

\author{Christopher~Rausch}
\affiliation{Fachbereich Physik and Landesforschungszentrum
OPTIMAS, Technische Universit\"at Kaiserslautern,
Erwin-Schr\"odinger-Stra{\ss}e 56, 67663 Kaiserslautern, Germany}

\author{Sebastian~Sch\"afer}
\affiliation{Fachbereich Physik and Landesforschungszentrum
OPTIMAS, Technische Universit\"at Kaiserslautern,
Erwin-Schr\"odinger-Stra{\ss}e 56, 67663 Kaiserslautern, Germany}

\author{Britta~Leven}
\affiliation{Fachbereich Physik and Landesforschungszentrum
OPTIMAS, Technische Universit\"at Kaiserslautern,
Erwin-Schr\"odinger-Stra{\ss}e 56, 67663 Kaiserslautern, Germany}

\author{Sang-Koog~Kim}

\affiliation{Research Center for Spin Dynamics \& Spin-Wave
Devices, Nanospinics Laboratory, Department of Materials Science
and Engineering, Seoul National University, Seoul 151-744,
Republic of Korea}

\author{Burkard~Hillebrands}

\affiliation{Fachbereich Physik and Landesforschungszentrum
OPTIMAS, Technische Universit\"at Kaiserslautern,
Erwin-Schr\"odinger-Stra{\ss}e 56, 67663 Kaiserslautern, Germany}

\date{\today}

\begin{abstract}
We present a new mechanism for spin-wave excitation using a pinned
domain wall which is forced to oscillate at its eigenfrequency and
radiates spin waves. The domain wall acts as a frequency doubler,
as the excited spin waves have twice the frequency of the domain
wall oscillation. The investigations have been carried out using
micromagnetic simulations and enable the determination of the main
characteristics of the excited spin-waves such as frequency,
wavelength, and velocity. This behavior is understood by the
oscillation in the perpendicular magnetization which shows two
anti-nodes oscillating out of phase with respect to each other.
\end{abstract}

\maketitle

The dynamic properties of ferromagnetic thin films have attracted
much attention recently. In particular, the possibility to create
logic circuits harvesting magnetic features such as domain walls
\cite{Allwood2005_1} and spin waves \cite{Schneider2008_1,
Schneider2008_2, Choi2007_1, Lee2008_1} is in the focus of
research activities. Thus, the excitation and propagation of spin
waves and their interaction with domain walls is relevant to this
research interest.

In this letter a new mechanism for the excitation and manipulation
of spin waves is presented. In case when a pinned domain wall is
excited by an external field with its eigenfrequency, a
"steady-state" oscillation forms with this eigenfrequency and a
distinct amplitude which is determined by the balance between
energy dissipation processes due to damping and the external
triggering by the applied field. The energy pumped into the system
by the external field leads not only to the compensation of the
damped oscillation, but also to the radiation of spin waves. It
will be shown that the domain wall itself oscillates with the
frequency of the externally applied field whereas the spin waves
of twice that frequency are allowed to propagate.

\begin{figure}
\begin{center}
\includegraphics[width=0.8\columnwidth]{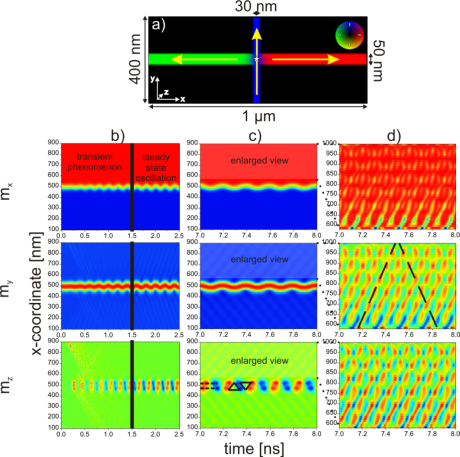}
\end{center}
\caption{\label{cut_at_y}(color online) a) Starting configuration.
The shape anisotropy fixes the magnetization in the arms in the
direction of the arms and forms a tail-to-tail domain wall. The
magnetization directions are displayed in the color code defined
by the color wheel shown in the upright corner. The denomination
of the axes is given in the bottom left corner. b-d) x-, y-, and
z-components of the magnetization along the x-axis of the system
mapped over time at the y-position shown in the cross area. b)
Temporal evolution of the system during the first 2.5\,ns. The
transient phenomenon in the first 1.5\,ns as well as the following
steady-state oscillation of the domain wall in the middle of the
structure can be clearly identified. c) Temporal evolution of the
system after the end of the transient phenomenon for a period of
1\,ns. The positions of the two anti-nodes in the z-component are
marked by the dashed lines on the left hand side. The triangular
shape of the domain wall oscillation is noteworthy and marked for
two different periods. d) Enlarged view of the arm region of the
structure for a duration of 1\,ns. The emission of spin waves from
the oscillating domain wall and their propagation is shown as
well. The slope of the straight lines given here is used to
calculate the spin-wave velocity.}
\end{figure}

To demonstrate the principle of a spin-wave frequency doubler,
micromagnetic simulations using the LLG-code \cite{LLG} were
performed. The used material is \Py\ and standard values for this
material (saturation magnetization $M_\mathrm{s}$=800\,G, exchange
constant $A_\mathrm{ex}$=1.05\,$\mu$erg/cm$^{3}$) were used. The
sample geometry is presented in Fig.~\ref{cut_at_y}(a). A
thickness of 5\,nm and cell sizes of 10\,nm for the x- and
y-direction and 5\,nm for the z-direction, respectively, have been
used. In this configuration the domain wall is pinned at the two
sides of the cross section area and therefore cannot move freely.
\cite{Petit2008_2} The starting configuration of the structure
exhibits a tail-to-tail domain wall and is presented in
Fig.~\ref{cut_at_y}(a).

The resonant excitation of the domain wall is one of the most
efficient means for creating spin waves. To determine this
resonance frequency, the complete structure has been excited by a
weak magnetic field pulse in x-direction. After performing a fast
Fourier transformation (hereafter referred to as FFT), the local
resonance frequency for each point of the structure can be
obtained. The resonance frequency for the domain wall has been
found to be 5\,GHz, and this frequency has been used as the
excitation frequency for the investigations presented in the
following. By exciting the structure with an external magnetic
field in x-direction (frequency 5\,GHz, amplitude of 10\,Oe), the
domain wall starts to oscillate and emits spin waves. The
amplitude of the external field is much smaller than the depinning
field necessary to drive the domain wall out of its original
position.

For a consistent study of this motion, the end of the transient
phenomenon, i.e.~the steady state of motion has to be reached. The
oscillating domain wall emits spin waves whose wavelength was
determined to be approximately 130\,nm. A detailed insight into
the observed phenomenon is given in Fig.~\ref{cut_at_y}. In this
figure, the temporal evolution of all three magnetization
components along the x-axis is shown for a fixed y-position in the
middle of the structure. The first column shows the first few
nanoseconds of the oscillation. The domain wall can be clearly
identified in the center of each panel. The externally applied
field needs some time to force the wall to oscillate. After this
transition which lasts approximately 1.5\,ns, a regular
oscillation of the wall can clearly be observed. The second column
of panels gives an enlarged view of the steady-state oscillation
over a duration of 1\,ns. It clearly shows that the wall
oscillates with a frequency of 5\,GHz.

The spin-wave emission in the wider arms can be observed best in
the y- and z-component (as defined in Fig.~\ref{cut_at_y}),
because these directions are perpendicular to the equilibrium
orientation of the magnetization, and are zero in the initial
state. In Fig.~\ref{cut_at_y}(b) the emission starts with the
first movement of the wall as can be seen from the z-component.
The m$_{z}$-component of the wall changes its sign during the
oscillation, as can be seen from the bright and dark areas in the
middle of the graph. As the amplitude of the spin waves is small
in comparison with the oscillation of the domain wall itself, the
spin waves are only weakly visible in Figs.~\ref{cut_at_y}(b) and
(c).

The velocity of the spin waves emitted from the wall can be
calculated using the slope of the straight lines indicated in
Fig.~\ref{cut_at_y}(d), yielding a value of 1.25\,$\mu$m/ns. The
interference pattern visible in Fig.~\ref{cut_at_y}(d) results
from spin waves created by the oscillating domain wall that
propagate along the arm until the end of the structure, where they
are reflected and overlap with the incoming waves.

The frequency of the propagating spin waves can be obtained by a
Fourier transform, which yields a value of 10\,GHz. Spin waves
with the excitation frequency of the external field of 5\,GHz can
not propagate since the corresponding wavevector is imaginary due
to the confinement of the spin waves by the stripe boundaries.

\begin{figure}
\begin{center}
\includegraphics[width=0.6\columnwidth]{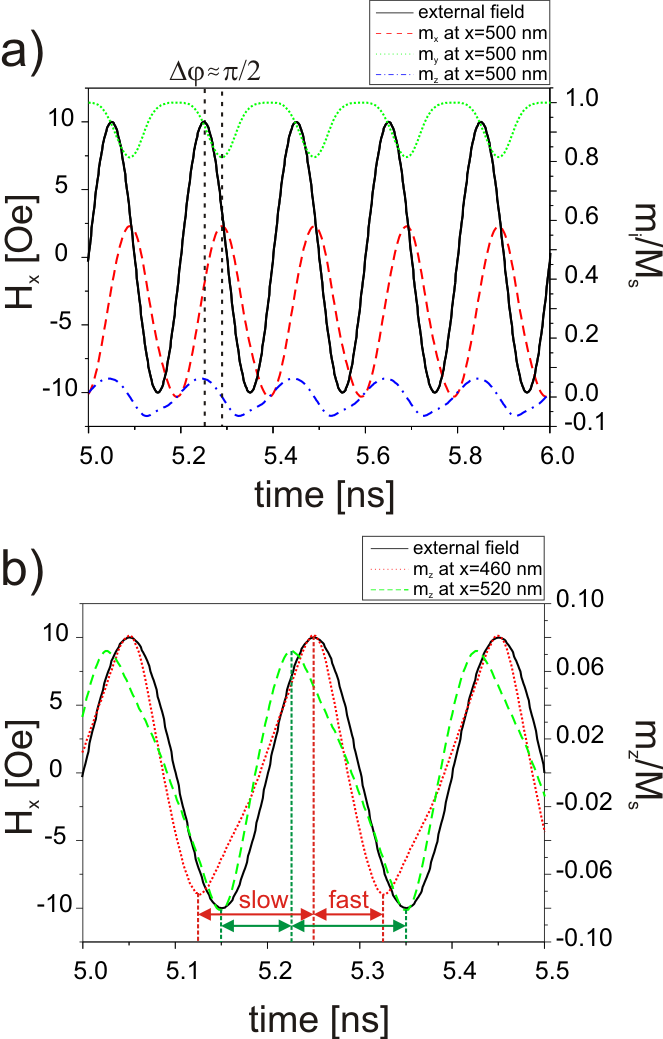}
\end{center}
\caption{\label{field_components}(color online) a) Oscillation of
the magnetization components at the point marked in
Fig.~\ref{cut_at_y}(a) in comparison to the externally applied
field. The amplitude of the external field is given on the left
hand side whereas the amplitude of the components is shown on the
right hand side. As expected for a driven oscillation the phase
shift between the driving field and the m$_{x}$- and
m$_{y}$-component, respectively, is nearly $\pi/2$. The
tilted-shaped m$_{z}$-component does not follow this phase
relation. b) m$_{z}$-component taken at the same y-position but at
the x-positions where the anti-nodes of the oscillation appear. It
can be clearly seen that one period of these oscillations shows a
slow increase and a fast decrease (dotted curve) or vice versa
(dashed curve). Note that the maxima of the dotted curve and the
minima of the dashed curve are in phase with the driving
oscillation.}
\end{figure}

As mentioned above, the spin-wave emission depends on the
oscillation in the m$_{z}$-component. This component shows two
anti-nodes as can be seen in Fig.~\ref{cut_at_y}(c). This slight
tilting of the magnetization out of the plane has already been
reported for transverse walls.\cite{Thomas2006_1} As expected for
a driven oscillation at the resonance frequency the m$_{x}$- and
m$_{y}$-components oscillate with a phase shift of nearly $\pi/2$
with respect to the externally applied field (see
Fig.~\ref{field_components}(a)). Unlike these two components, the
m$_{z}$-component does not follow this phase relation and shows a
tilted oscillation curve which is partially in phase with the
external field. This behavior is expected as the domain wall has
to be driven by the externally applied field and, therefore, its
phase shift is fixed to nearly $\pi/2$. The domain wall can be
monitored best in the m$_{x}$- and m$_{y}$-component as in these
components it appears as the region with m$_{x}$=0 and m$_{y}$=1,
respectively. Since the total magnetization remains constant, the
motion of the domain wall, i.e.~a moving area with m$_{y}$=1,
causes the other two components to be zero at this position. In
particular, this means that the m$_{z}$-component is zero (as the
m$_{x}$-component is zero by definition at the position of the
domain wall) and explains why this component is in phase with the
external field.

As can be seen from Fig.~\ref{field_components}(b), the waveforms
of the oscillations at the positions, where the anti-nodes of the
oscillation occur (i.e.~at a position of 460\,nm and 520\,nm,
respectively), are asymmetric in deviation from the sinusoidal
waveform of the exciting external field and cause second harmonics
generation in the spin-wave excitation.

\begin{figure}
\begin{center}
\includegraphics[width=0.6\columnwidth]{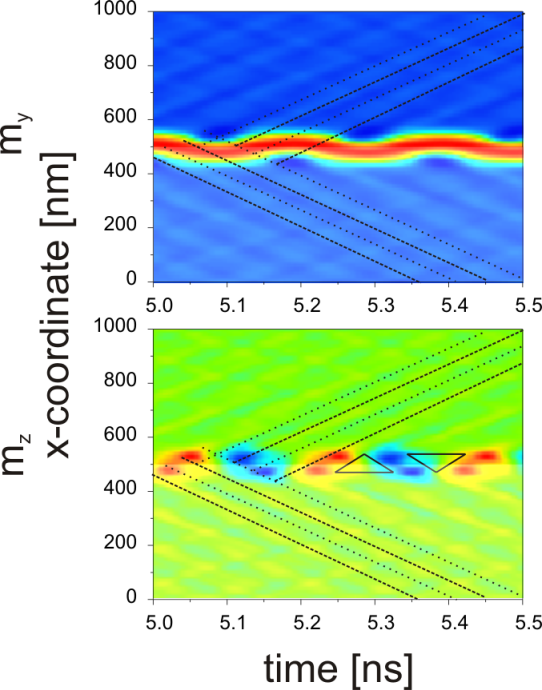}
\end{center}
\caption{\label{SW_radiation}(color online) Enlarged view of y-
and z-components of the magnetization as presented in
Fig.~\ref{cut_at_y}. In this case an amplitude of 40\,Oe has been
taken to clarify the domain wall movement as well as the spin-wave
excitation. The lower parts of the figures were digitally enhanced
for better visualization. The triangular shape of a full period of
the oscillation in the z-component is clearly visible and marked
for one period. Red areas show a positive value of this component,
blue areas a negative one. The position of the two anti-nodes can
be identified as well as the fast and slow switching leading to
the triangular appearance. The excited spin waves are marked by
the straight lines as guides to the eye. The dashed lines in the
z-component mark maxima, the dotted lines minima. By comparing the
two components and the course of the lines, the spin waves are
excited by the oscillation of the anti-nodes as marked by the
lines. The fact that two of these points exist and oscillate
explains the frequency doubling.}
\end{figure}

Fig.~\ref{SW_radiation} presents an enlarged view of
Fig.~\ref{cut_at_y}. The characteristic triangular shape of the
z-component as presented in Fig.~\ref{cut_at_y}(c) can be
identified, resulting from the different velocities of the
switching of the m$_{z}$-component from positive to negative and
vice versa. The frequency of this oscillation is still 5\,GHz, as
one red and blue area represents half a period each. The frequency
doubling is clearly visible by the four beams radiated to both
sides within one period (i.e. two maxima (dashed) and two minima
(dotted)) and shows that the oscillation of the m$_{z}$-component
is essential for the excitation of the spin waves. The origin of
these spin waves are the two oscillating anti-nodes of the
oscillation, as can be seen from Fig.~\ref{SW_radiation}. The
emission of two maxima in the bottom part of the z-component and
one maximum in the top part in the case of positive anti-nodes and
one minimum in the bottom part and two maxima in the top part for
negative anti-nodes within one period can be identified. The
oscillation of both of these points excites a new spin wave
wavefront when switching from the positive m$_{z}$-component to
the negative and vice versa.

Thus, all three components oscillate with the frequency of the
external field. As a result of the two anti-nodes in the
m$_{z}$-component which excite the spin waves by their
oscillation, this component oscillates effectively with twice the
driving frequency and allows for the excitation and frequency
doubling of the waves.

In summary, dynamic micromagnetic simulations reveal the
possibility to create spin waves by the excitation of a pinned
domain wall with an external magnetic field. The pinning is
necessary to drive the excited domain wall back to its equilibrium
position. The properties of the created propagating spin waves
such as frequency, wavelength, and velocity can be determined from
the simulation data. The frequency doubling of the excited spin
waves with respect to the externally applied field can be
understood by the oscillation of the m$_{z}$-component. As this
component shows two anti-nodes which are oscillating non-uniformly
but each of them with the given frequency, spin-wave wavefronts
are created by both of these oscillations and frequency-doubled as
two oscillations contribute to it.

Financial Support by the DFG within the Priority Programme 1133
"Ultrafast magnetization processes" is gratefully acknowledged.
S.-K. Kim is supported by Creative Research Initiatives (Research
Center for Spin Dynamics and Spin-Wave Devices) of MEST/KOSEF. The
authors would like to thank Doroth\'{e}e Petit from the Imperial
College London for fruitful discussion and Simon Trudel for
careful proof-reading.

\newpage

\begin{thebibliography}{10}
\bibitem{Allwood2005_1} D.~A.~Allwood, G.~Xiong, C.~C.~Faulkner, D.~Atkinson,
D.~Petit, and R.~P.~Cowburn, Science {\bf 309}, 1688 (2005).
\bibitem{Schneider2008_1} T.~Schneider, A.~A.~Serga, B.~Leven, B.~Hillebrands, R.~L.~Stamps, and M.~P.~Kostylev, Appl. Phys. Lett. {\bf 92},
022505 (2008).
\bibitem{Schneider2008_2} T.~Schneider, M.~P.~Kostylev,
A.~A.~Serga, and B.~Hillebrands, J. Nanoelectron. Optoelectron.
{\bf 3}(1), 69 (2008).
\bibitem{Choi2007_1} S.~Choi, K.-S.~Lee, K. Yu. Guslienko, and S.-K.~Kim, Phys. Rev. Lett. {\bf 98},
087205 (2007).
\bibitem{Lee2008_1} K.-S.~Lee and S.-K.~Kim, J. Appl. Phys. {\bf 104},
053909 (2008).
\bibitem{LLG} M.R.~Scheinfein, LLG Micromagnetic Simulator\texttrademark,
http://llgmicro.home.mindspring.com.
\bibitem{Petit2008_2} D.~Petit, A.-V.~Jausovec, H.~T.~Zeng, E.~Lewis, L.~O’Brien, D.~Read, and R.~P.~Cowburn, Appl. Phys. Lett. {\bf 93},
163108 (2008).
\bibitem{Thomas2006_1} L.~Thomas, M.~Hayashi, X.~Jiang, R.~Moriya,
C.~Rettner, and S.~S.~P.~Parkin, Nature {\bf 443}, 197 (2006).
\end{thebibliography}
\end{document}